\documentclass[12pt]{article}

\usepackage[margin=1in]{geometry}
\usepackage{setspace}
\usepackage{graphicx}
\usepackage{refstyle}
\usepackage{caption}
\usepackage{subcaption}
\usepackage{multirow}
\usepackage{authblk}
\usepackage{amsmath,amssymb}

\title{Substitutional Doping of Symmetrical Small Fullerene Dimers}
\author[1,a]{Sandeep Kaur}
\author[1,b]{Amrish Sharma}
\author[2,c]{Hitesh Sharma}
\author[3,d]{Shobhna Dhiman}
\author[4,e]{Isha Mudahar}
\affil[1]{Department of Physics, Punjabi University, Patiala, Punjab, India}
\affil[2]{Department of Applied Sciences, IKG Punjab Technical University, Amritsar, Punjab, India}
\affil[3]{Department of Applied Sciences, Punjab Engineering College, Chandigarh, India}
\affil[4]{Department of Basic and Applied Sciences, Punjabi University, Patiala, Punjab, India}
\affil[a]{sandeep\_rs16@pbi.ac.in}
\affil[b]{amrish99@gmail.com}
\affil[c]{hitesh@ptu.ac.in}
\affil[d]{shobhnadhiman1962@gmail.com}
\affil[e]{dr.ishamudahar@gmail.com}

\begin{document}

\date{}
\maketitle
\doublespace

\begin{abstract}
Magnetic carbon nano-structures have potential applications in the field of spintronics as they exhibit valuable magnetic properties. 
Symmetrically sized small fullerene dimers are substitutional doped with nitrogen (electron rich) and boron (electron deficient) atoms to visualize the effect on their magnetic properties. 
Interaction energies suggests that the resultant dimer structures are energetically favourable and hence can be formed experimentally. 
There is significant change in the total magnetic moment of dimers of the order of 0.5 $\mu_{B}$ after the substitution of C atoms with N and B, which can also be seen in the change of density of states. 
The HOMO-LUMO gaps of spin up and spin down electronic states have finite energy difference which confirm their magnetic behaviour, whereas for non-magnetic doped dimers, the HOMO-LUMO gaps for spin up and down states are degenerate. 
The optical properties show that the dimers behave as optical semiconductors and are useful in optoelectronic devices.
The induced magnetism in these dimers makes them fascinating nanocarbon magnetic materials.
\end{abstract}

\section{Introduction}

The discovery of $C_{60}$, among fullerene family has triggered an interest in the field of carbon nano-structured materials due to their fascinating physical and electronic properties\cite{kroto, Wkroto, kratschmer}. 
An intensive research has been initiated for large and small fullerene derivatives after the large-scale synthesis of $C_{60}$~\cite{kroto, kratschmer, diederich}.
Small fullerenes have possible usage in the field of nano-electronics, spin-electronics, molecular devices, superconducting devices, drug delivery and energy storage owing to their unique physical and chemical properties~\cite{Wkroto, piskoti, an, celaya, breda}. 
$C_{20}$ is the smallest fullerene cage which was synthesized using gas-phase debromination\cite{prinzbach} and was found to be less stable kinetically than higher fullerenes such as $C_{36}$ or $C_{60}$ due to its strong curvature~\cite{podlivaev, Apodlivaev}. 
The carbon cages like $C_{32}$, $C_{44}$ and $C_{50}$ are important because of their large ionization potentials and band gaps~\cite{tendero}.
The spectrum of 11.2 ${\mu}$m unidentified infrared band (UIR) indicates that $C_{24}$ fullerene cage can be used as a carrier which acts as a useful probe for astrophysical environment~\cite{bernstein}.

The surface of the fullerenes, depending on its inter and intra curvature, determine their stability and electronic properties. 
The ability of the surface to react with other objects strongly depends on its ability to form chemical bonds. 
In carbon networks, the substitution of N and B is strongly favourable as they bracket carbon in periodic table.
The substitutional doping of N and B in fullerene cage structures leads to increase in their static polarizabilities\cite{sedel}.
The first N doped derivatives of $C_{60}$ and $C_{70}$ were synthesized using contact-arc vaporization of graphite in the vicinity of gaseous $N_2$ and $NH_3$~\cite{pradeep}. 
The isolation of series of N-doped derivatives stimulated their theoretical studies further~\cite{Ryu, otero}. 
The modification in chemical composition of small fullerenes can increase their stability~\cite{strout}. 
The substitutional doping of $C_{20}$ with transition metals, B and N atoms alters the charge distribution of cage, stability, polarity and electronic structure of small fullerenes~\cite{rad, amiri}. 
Substitution of N atoms at hetero-position in $C_{20}$ cage induce Stone-Wales defect which further depend on number and relative position of N atoms in $C_{20}$ cage~\cite{katin}. 
Though $C_{36}$ cage like $C_{60}$ can donate or gain maximum of 6 electrons but it has different properties\cite{fowler}.
The substitution of N and B in $C_{36}$ cage alters the electronic properties and have tendency to make polymer structures which have higher stability than their carbon counterparts\cite{maryam}.
The electronic structure of $C_{50}$ isomers substituted by B and N form acceptor/donor pairs, which could be useful in molecular rectification~\cite{viani}.

Small fullerene cages can also form dimer structures, as they have an ability to form inter-cage bonds~\cite{kaur}. 
The dimerization of $C_{60}$ fullerenes can occur in different phases like peanut, dumb-bell and capped nanotubes, which has been confirmed experimentally~\cite{GWwang, Nkaur, Kkomatsu}.
The study on polarizability of $C_{60}$ dimers have shown that the effect of exaltation of polarizability occur due to interaction of $\pi$-electronic systems of fullerene cores\cite{sabirov}.
Similar to $C_{60}$, the exaltation of polarizability also occur when small fullerene cages are united to form dimers and is typical property for all the members of fullerene family ($C_{20}$, $C_{24}$, $C_{30}$, $C_{36}$, $C_{50}$ and $C_{70}$)\cite{denis}.
The determination of molecular polarizability using Thole's model have also shown that there is enhancement in the polarizabilities of $C_{60}$ oligomers\cite{swart}.
The effect of exaltation in polarizabilities of fullerene dimers is a common property and is independent on their chemical nature and topological features\cite{alina}.

The optical response of carbon allotropes have been tuned to make them useful for UV light protection, light-to-energy conversion etc\cite{bhattacharya, NBsingh}. 
$C_{60}$ cage possess a low static dielectric function having magnitude of 4.4 which is similar to that of diamond\cite{ching}.  
The optical gap of $C_{24}$ fullerene can be tuned with doping from visible to UV region\cite{paul}.
The strong absorption occur in UV region for $C_{28}$ fullerene and their low values of reflectivity are useful for hybrid solar cells\cite{Dpaul}.
The optical properties of both pure and doped dimers can be a very significant study.

In our previous work, we have reported a systematic study on symmetrical and asymmetrical sized small fullerene dimers, which are likely to be formed experimentally~\cite{kaur}. 
Our results have shown that the formation of small fullerene dimers strongly depend on the type of inter-connecting bonds between two cages. 
The substitution of carbon atoms with N and B atoms at junction of two cages can affect the structural, electronic, magnetic and optical properties of these dimer structures.
In the present work, the substitutional doping of N and B at connecting bonds of symmetrically sized small fullerene dimers and their structural, electronic, magnetic and optical properties are studied. 
Our results show the possibility of the formation of such dimers with significant induced magnetic moment which can further be used in nanocarbon magnetic materials.

\section{Computational Details}

The first principle calculations based on density functional theory were performed using Spanish Initiative for Electronic Simulation with Thousands of Atoms (SIESTA) computational code~\cite{junquera}. 
The double-zeta polarized basis set combined with Perdew, Burkey and Ernzerhof (PBE) functional was used for the geometry optimization~\cite{perdew}.
The core electrons were described by using Kleinman-Bylander form of non-local norm conserving pseudo potentials~\cite{kleinman} whereas numerical pseudoatomic orbitals of the Sankey-Niklewski type~\cite{sankey} were used to represent the valence electrons. 
The size of pseudoatomic orbitals is defined by energy shift parameter ~ 325 meV. 
The fineness of a finite grid is defined by mesh cut-off value i.e. 175 Ry. 
In order to obtain ground state properties, minimization of total energy of the system has been executed by relaxing the residual forces of the system up to 0.004 eV/{\AA}. 
The energy eigen values have been plotted in order to get density of states (DOS) spectra.

The test calculations have been performed on small fullerene cages to check the accuracy of parameters used in our calculations. 
We found that calculated values of average bond lengths, average diameters and HOMO-LUMO gaps are in good agreement with known theoretical and experimental results~\cite{piskoti, feyereisen}. 
The input parameters have been checked since our group has already studied carbon based systems~\cite{garg, hitesh, amrish, skaur, asharma}.

\section{Results and Discussion}
There are four different modes through which fullerene cages can connect to form dimer structures\cite{Anafcheh, Ma, kaur}. 
These four modes are: $[1+1]$ point mode, $[2+2]$ side mode, $[5+5]$ face mode between two pentagon rings and $[6+6]$ face mode between two hexagon rings ~\cite{kaur, Ma}. 
These modes have been studied in endohedral fullerene dimers Na@$C_{60}$-$C_{60}$@F and have larger bond energies as compared to $\pi-(C_{60})_{2}$\cite{Ma}. 
The most favourable bonding between $C_{60}$ molecules is [2+2] cycloaddition through (i) dimerization via nucleophillic addition and (ii) dimerization via coupling through an electron transfer step\cite{segura}. 
The endohedral complexes of $C_{60}$ dimers have higher stability in [1+1] mode whereas hyperpolarizabilities are higher for [5+5] mode\cite{Ma}.

\begin{figure}
\centering
\includegraphics[width=5in]{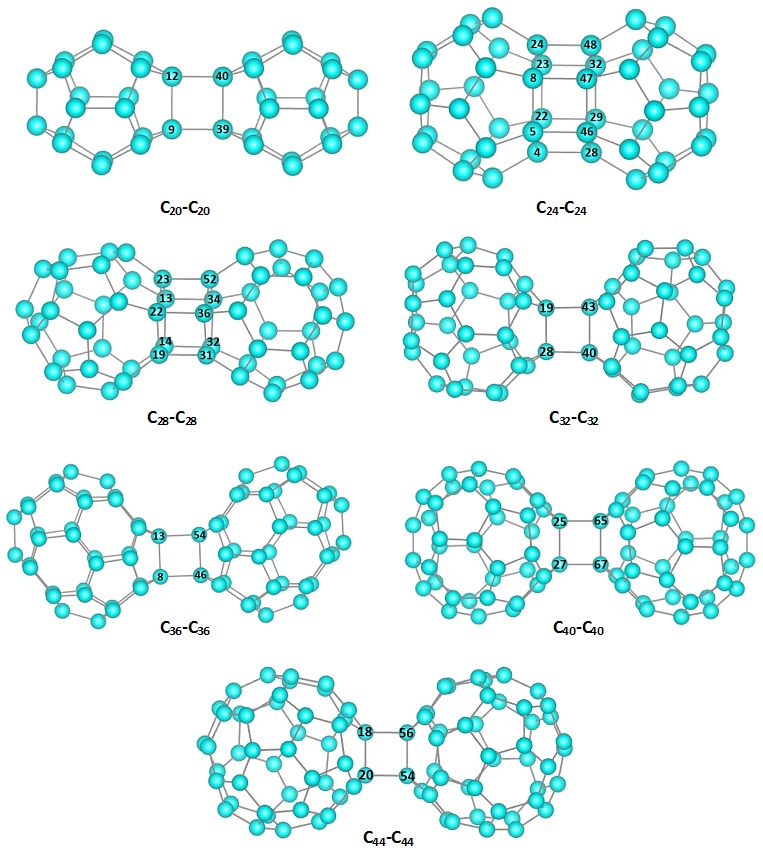}
\caption{Doping positions for all symmetrically sized dimers.}
\label{fig:doping}
\end{figure}

In our previous work\cite{kaur}, we have considered all the four modes to find the most stable configuration for small fullerene dimers. $C_{24}-C_{24}$ and $C_{28}-C_{28}$ dimers have higher stability in $[6+6]$ and $[5+5]$ modes respectively, whereas for $C_{20}-C_{20}, C_{32}-C_{32}, C_{36}-C_{36}, C_{40}-C_{40}$ and $C_{44}-C_{44}$ dimers, $[2+2]$ side mode is most favourable\cite{kaur}. 
These modes have been considered to visualize the effect of substitutional doping of small fullerene dimers. 
In these most stable configurations, the two C atoms at connecting bond of dimer structures are replaced with one nitrogen (N) and one boron (B) atoms, which alter the electronic as well as magnetic properties due to donor/acceptor tendency of N and B respectively.
The doping positions of N and B are shown in Figure \ref{fig:doping}. 
In all dimer structures, positions shown on the left cage are doping positions for N atom, whereas for B atom doping positions are shown on the right cage.
The stability for all the doped dimer structures depend on interaction energy $(E_{int})$ which can be calculated using following expression
\begin{equation}
E_{int} = E_{C_{n}N-C_{n}B}-E_{C_{n}N}-E_{C_{n}B}-E_{CC}
\end{equation}
where $E_{C_{n}N-C_{n}B}$ is total energy of doped dimer structure, $E_{C_{n}N}$ is total energy of fullerene cage doped with N, $E_{C_{n}B}$ is total energy of fullerene cage doped with B and $E_{CC}$ is counterpose correction term which is considered here to remove basis set superposition error (BSSE) in order to get more realistic values of interaction energy. 
The interaction energies for most stable structures of pure dimers~\cite{kaur} and their doped derivatives are tabulated in Table 1. 
The negative sign indicates the dimer structures are energetically favourable and are likely to be formed experimentally. 
The more negative number shows the higher stability of doped dimer.

The chemical hardness ($\eta$) and electronegativity ($\chi_{m}$) for these doped dimers are calculated by using HOMO-LUMO energies and are given by following equations:
\begin{equation}
\eta = \dfrac{E_{LUMO}-E_{HOMO}}{2}
\end{equation}
and
\begin{equation}
\chi_{m} = - \dfrac{E_{HOMO}+E_{LUMO}}{2}
\end{equation}
respectively. 
The calculated parameters are tabulated in Table 1.
The HOMO and LUMO energy levels play important role in chemical stability\cite{adel}. 
The lower HOMO-LUMO gap results in more intra-molecular charge transfer and hence more chemical activity of molecule. 
As compared to other B and N doped carbon nanomaterials like sumanene\cite{adel,vanani}, these doped dimers have higher chemical activity due to the lower values of $\eta$. 
The relative stability of a molecule can be determined by its chemical hardness and the stability is directly related to higher values of hardness\cite{adel, vanani}. 
The calculated values of $\eta$ for these doped dimers are lower than that for B and N doped sumanene. 
The $\chi_{m}$ values for these structures are comparable to B doped sumanene\cite{adel} but higher than their N doped derivatives\cite{vanani}.

The functions related to optical properties of the fullerenes are deduced from frequency-dependent complex dielectric function, $\varepsilon$ = $\varepsilon_{1}(\omega)$ + $i\varepsilon_{2}(\omega)$, where $\varepsilon_{1}(\omega)$ and $\varepsilon_{2}(\omega)$ are the real and imaginary parts of dielectric function. 
Fermi's golden rule was used to obtain the imaginary part of dielectric function due to interband transitions using Kramers-Kronig relations and the real part was then derived from imaginary part.
The real and imaginary part of dielectric function were used to calculate other functions such as absorption coefficient ($\alpha(\omega)$), the optical conductivity ($\sigma(\omega)$), the reflectivity (R($\omega)$) and the refractive index (n($\omega$)) as follows:
\begin{equation}
\alpha(\omega) = \sqrt{2}\omega[\sqrt{\varepsilon_{1}^{2}(\omega)+\varepsilon_{2}^{2}(\omega)}-\varepsilon_{1}(\omega)]^{r/2}
\end{equation}
\begin{equation}
\sigma(\omega) = -i \dfrac{\omega}{4\Pi}|\varepsilon(\omega)-1|
\end{equation}
\begin{equation}
R(\omega) = [ \dfrac{\sqrt{\varepsilon_{1}(\omega)+i\varepsilon_{2}(\omega)}-1}{\sqrt{\varepsilon_{1}(\omega)+i\varepsilon_{2}(\omega)}+1}]^{2}
\end{equation}
\begin{equation}
n(\omega) = [\sqrt{\varepsilon_{1}^{2}(\omega)+\varepsilon_{2}^{2}(\omega)}+\varepsilon_{1}(\omega)]^{2}/\sqrt{2} 
\end{equation}
where $\omega$ is the frequency of the electromagnetic radiation.

\subsection{Structural Properties}
The pure $C_{20}-C_{20}$ dimer is most stable in [2+2] mode, which indicates that there are two possible way of N and B doping on the two C-C connecting bonds~\cite{kaur}. 
Among two isomers, the dimer in which N and B are substituted at (9,39) position is more stable in comparison to the dimer with (N, B) doping at (12,40) position, which has relative energy difference of 0.16 eV. 
The doped $C_{20}$ dimer has higher stability than pure $C_{20}$ dimer as a result of more negative value of interaction energy. 
When N and B atoms are substituted at (9,39) position, the dimer attains larger interaction energy and hence higher stability as compared to (12,40) position of N and B. 
The chemical hardness for (9,39) have been found to be higher in comparison to (12,40) indicating their higher stability. 
The electronegativity ($\chi_{m}$) has found to be independent of the position of doped atoms.
The connecting bond length of C-C bond in dimer is decreased from $1.55$ {\AA}~\cite{kaur} to $1.36$ {\AA} after doping. 
The bond length of N - B bond is $1.40$ {\AA} which is slightly lower than B-N bond of hexagonal boron nitride ($1.44$ {\AA}). 
The average diameter of doped dimer decreases to $4.06$ {\AA} from $4.17$ {\AA}~\cite{kaur}. 
The doping of $C_{20}$ dimer at junction causes opening of fullerene cages with formation of tube like structure.

$C_{23}N-C_{23}B$ dimer has six isomers due to six connecting bonds in most stable [6+6] structure. 
The six positions for substitutional doping of (N,B) are (4,28), (5,46), (8,47), (22,29), (23,32) and (24,48). 
Table 1 shows that the stability of dimer decreases after its substitutional doping. Among doped isomers, (22,29) position of (N,B) has highest stability, whereas other isomers have finite energy difference with respect to (w.r.t.) (22,29). 
The $\eta$ and $\chi_{m}$ of most stable isomer structures have shown similar value with very small variation. The bond length of N - B connecting bond is 1.68 {\AA}, which is larger than N - B bond (1.44 {\AA}) of hexagonal boron nitride, whereas C - C bond length is 1.58 {\AA} which is comparable to connecting bond length of pure $C_{24}$ dimer ~\cite{kaur}. The average diameter of both cages remains almost same with doping of dimer structure.

$C_{27}N-C_{27}B$ dimer has five isomers having doping positions of (N,B) at (13,34), (14,32), (19,31), (22,36) and (23,52). Interaction energy shows that with substitutional doping of $C_{28}$ dimer, the stability increases. 
Among its five isomers, the most stable doping position is (13,34), whereas other positions have relative energy difference w.r.t. most stable configuration. (13,34), (14,32) and (22,36) have higher chemical hardness values as compared to other isomers while electronegativity is higher for (N,B) position at (23,52). The connecting bond length of N - B and C-C bond in most stable dimer is 1.73 {\AA} and 1.59 {\AA} respectively. 
The average diameter of both the cages remain same after substitution of foreign atoms.

For $C_{31}N-C_{31}B$, $C_{35}N-C_{35}B$, $C_{39}N-C_{39}B$ and $C_{43}N-C_{43}B$ dimers, there are two isomers, as their [2+2] mode is most stable. The stability of dimer structures increases after their doping, except for $C_{36}$ dimer which shows decrease in stability w.r.t. doped structure. In $C_{31}N-C_{31}B$ dimer, the most favourable position for (N,B) is (19,43), while relative energy difference for (28,40) position is 1.29 eV. The $\eta$ and $\chi_{m}$ values are higher for (28,40) position of (N,B). For $C_{35}N-C_{35}B$ dimer, (8,46) position is most stable, whereas $E_r$ for (13,54) position is 0.74 eV. The two positions (25,65) and (27,67) in $C_{39}N-C_{39}B$ dimer are isoenergetic, whereas in $C_{43}N-C_{43}B$ dimer (18,56) position is the most favourable one and $E_{r}$ for (20,54) position is 0.52 eV. The chemical hardness and electronegativity for $C_{35}N-C_{35}B$, $C_{39}N-C_{39}B$ and $C_{43}N-C_{43}B$ dimers are equal for their isomers. The connecting bond length of N-B bond is ~1.55 {\AA} and C-C connecting bond length is ~1.67 {\AA} for all these dimers. The average diameter varies for doped dimers as compared to their pure counterparts.

Hence substitution of N and B at junction of small fullerene dimer cages leads to change in stability as well as structural parameters. The stability of dimers increases with their doping except for $C_{24}$ and $C_{36}$, which are more stable in their pure form. The doping in $C_{20}$ and $C_{40}$ cage leads to opening of the fullerene cages at the junction of dimers.

\subsection{Electronic and Magnetic Properties}
\begin{figure}
\centering
\includegraphics[width=5in]{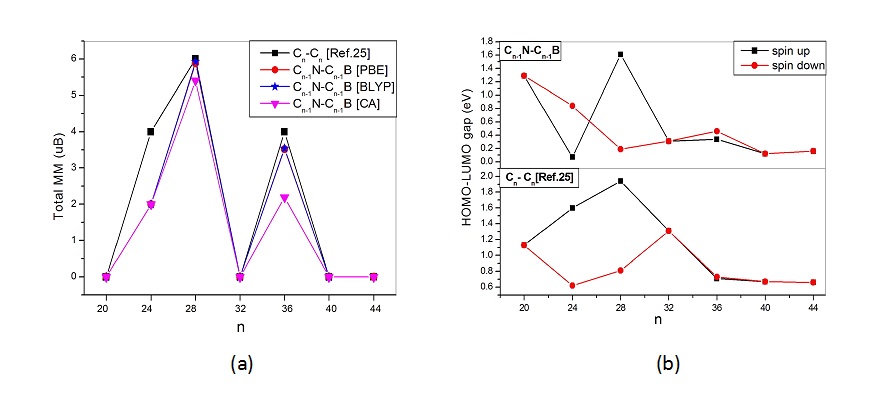}
\caption{Total Magnetic Moments and HOMO-LUMO gaps for symmetrical small fullerene dimers.}
\label{fig:Tgap}
\end{figure}

\begin{figure}
\centering
\includegraphics[width=6in]{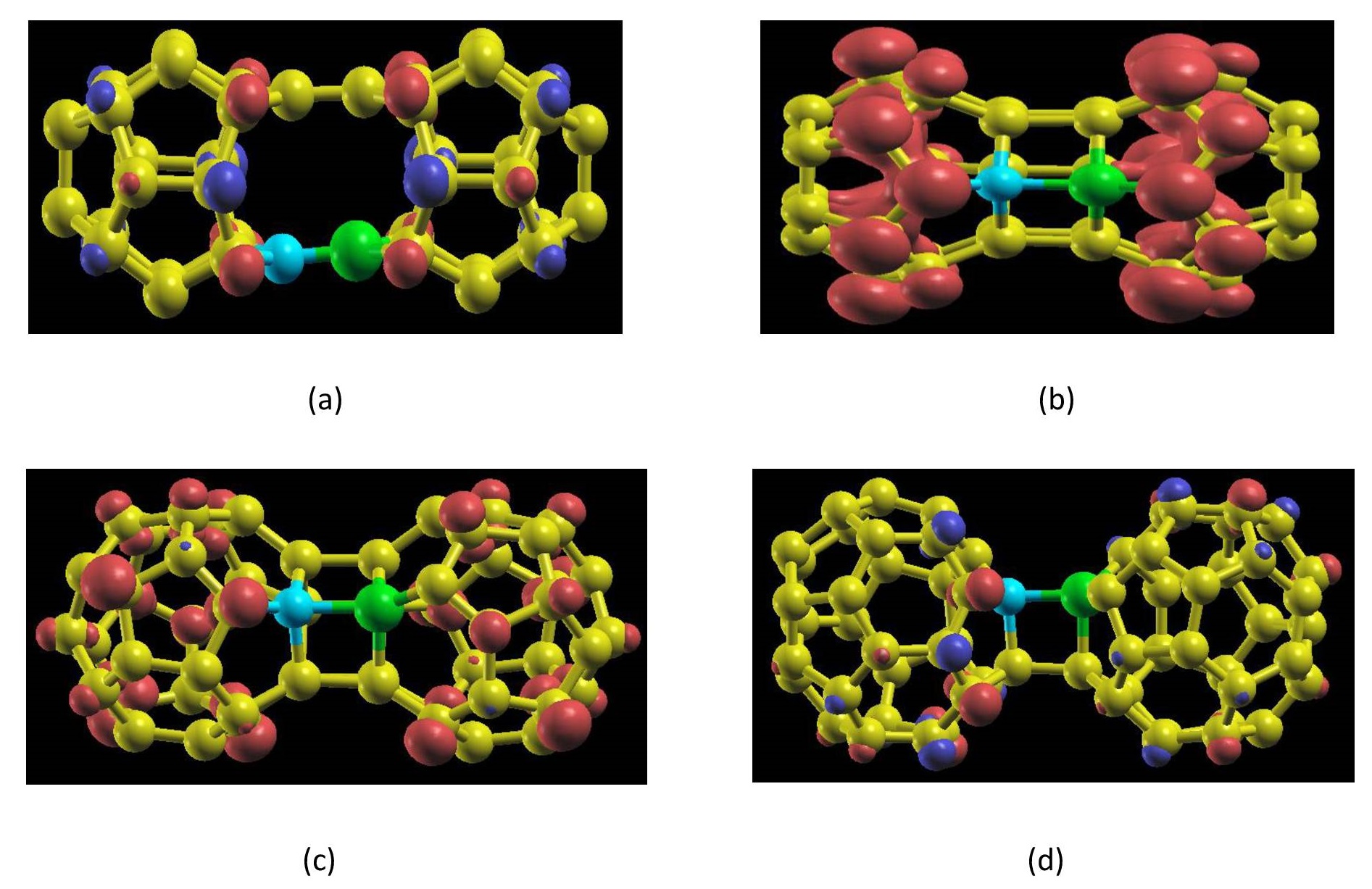}
\caption{Local Magnetic Moments for a) $C_{19}N-C_{19}B$, b)$C_{23}N-C_{23}B$, c) $C_{27}N-C_{27}B$ and $C_{31}N-C_{31}B$. The yellow, green and sky blue balls refer to carbon, boron and nitrogen atoms.}
\label{fig:localmm}
\end{figure}

\begin{figure}
\centering
\includegraphics[width=6in]{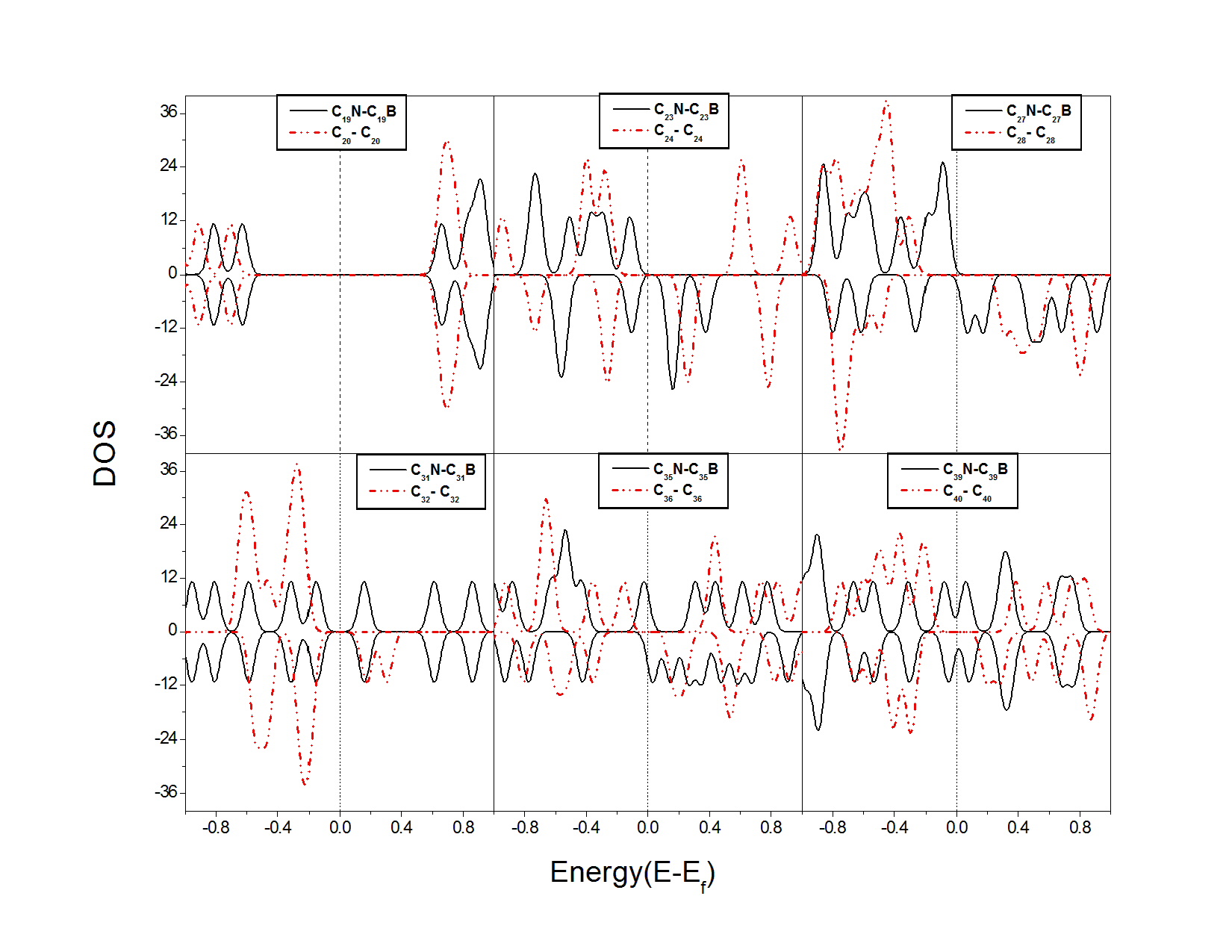}
\caption{Density of states for doped symmetrical small fullerene dimers. The plots are compared with DOS for pure dimers [Ref. 25].}
\label{fig:dos}
\end{figure}

The spin polarized calculations have been performed to study the magnetic properties of substitutionally doped small fullerene dimers. The HOMO-LUMO gaps for spin up and spin down electronic states, total magnetic moments (TMM), local magnetic moments (LMM), Muliken charge analysis and density of states (DOS) have been studied and are shown in Figure \ref{fig:Tgap} and \ref{fig:dos}, whereas localized MMs for some particular cases are plotted in Figure \ref{fig:localmm}. In our earlier work we have investigated the pristine dimer structures systematically\cite{kaur}. The electronic and magnetic properties of the doped dimers presented in this paper have also been compared with the pristine dimers for better understanding of the change in the properties. We have also calculated the TMMs in different correlation fields as shown in Figure \ref{fig:Tgap}. The TMM has been found to be same with PBE and BLYP exchange correlation functional. However, the LDA (Local Density Approximation) approximation considered through CA (Ceperley Alder) underestimates the TMMs due to the treatment of exchange correlation in Local Density field.

The $C_{20}$ dimer and N,B substituted $C_{20}$ dimer both are non-magnetic. 
Figure \ref{fig:localmm} and \ref{fig:dos} show localized MMs and DOS respectively for $C_{19}N-C_{19}B$ dimer, which confirm its non-magnetic nature owing to its equal spin up and spin down charge density.
The HOMO-LUMO gaps for spin up and spin down states are equal as all the electrons are paired. Muliken charge analysis shows the electron donor and acceptor behaviour between N and B atom where N donates 0.56 electrons to B, whereas additional charge is gained by B atom from the surrounding C atoms (Table 2). 

Pure $C_{24}$ is magnetic in nature having total MM of 4 $\mu_B$ ~\cite{kaur}. The substitution of C with N and B at junction of two cages in $C_{23}N-C_{23}B$ dimer structure results in decrease in total MM from 4 $\mu_B$ to 2 $\mu_B$. TMM for all five isomers is found to be identical. The localized MM shows that the major contribution towards total MM comes from first and second nearest neighbours (NN) near connecting bonds. The localized MM also confirm the ferromagnetic nature of interaction between two cages. The plotted spin up and down electronic densities are unequal which shows the magnetic nature of the dimer. The DOS shows significant change near the Fermi level after doping, which indicates the increase of conductivity of dimer with doping and it become half-metallic which also shows finite energy difference in its spin up and down HOMO-LUMO gap. The unpaired electrons in highest occupied orbital leads to the magnetic nature of dimer structure. There is a transfer of 0.4 electrons from N to B atom, whereas additional charge on B atom is because of charge transfer from the surrounding C atoms.

$C_{27}N-C_{27}B$ and $C_{35}N-C_{35}B$ both show decrease in the magnitude of TMM w.r.t. their pure structures. For $C_{27}N-C_{27}B$ dimer it reduces from 6 $\mu_B$ to 5.87 $\mu_B$, whereas for $C_{35}N-C_{35}B$ it decreases from 4 $\mu_B$ to 3.52 $\mu_B$. The localized MM shows that the major contribution towards total MM comes from first and third nearest neighbours from the connecting bonds. The reduction in the total MM w.r.t. doped dimer may be due to the charge transfer between N and B atoms. As B is an electron deficient element, so it gains charge from N and surrounding C atoms, which leads to quenching in local MM from these C sites, resulting in the reduction of magnetism (Table 2). Both the dimer structures have ferromagnetic interaction as spin up charge density is higher than spin down charge density. DOS plots also confirm the same magnetic behaviour. The conductivity of doped dimer increases with doping due to the modification of electronic states near Fermi level. The finite energy difference between HOMO-LUMO gaps of spin up and spin down states is consistent with the electronic DOS and explain the magnetic nature of these dimers.

It is important to highlight that the pristine dimers of $C_{32}$, $C_{40}$ and $C_{44}$ are non-magnetic\cite{kaur} and their state remains unaltered even after doping with N and B atoms. The calculation of on site local MM further confirms the non-magnetic nature of the doped dimers. The electronic DOS of these dimers shows significant change in the density distribution near the Fermi level w.r.t. the pristine dimers, which shows decrease in HOMO-LUMO gap. $C_{40}$ dimer becomes metallic with substitutional doping as more empty states are created near the Fermi level due to redistribution and polarization of electrons. The spin up and spin down HOMO-LUMO gaps are equal for all these dimers which further confirm their non-magnetic nature. Muliken charge analysis show that N behaves as electron donor whereas B atom accept electrons resulting in transfer of charge between the two cages.

Substitution of N and B in symmetrical dimers alter their electronic and magnetic properties significantly w.r.t. their pristine configuration. The change in the magnetic properties can be explained on the basis of change in localized MM of C atoms near the connecting bonds on doping. The substitution with N and B leads to decrease in magnetic moment of $C_{24}$, $C_{28}$ and $C_{36}$ dimers, while $C_{20}$, $C_{32}$, $C_{40}$ and $C_{44}$ dimers do not show any change in their non-magnetic nature. However, the conductivity of these dimers increase due to decrease in HOMO-LUMO gap and charge transfer/redistribution.

\subsection{Optical Properties}

\begin{figure}
\centering
\includegraphics[width=6in]{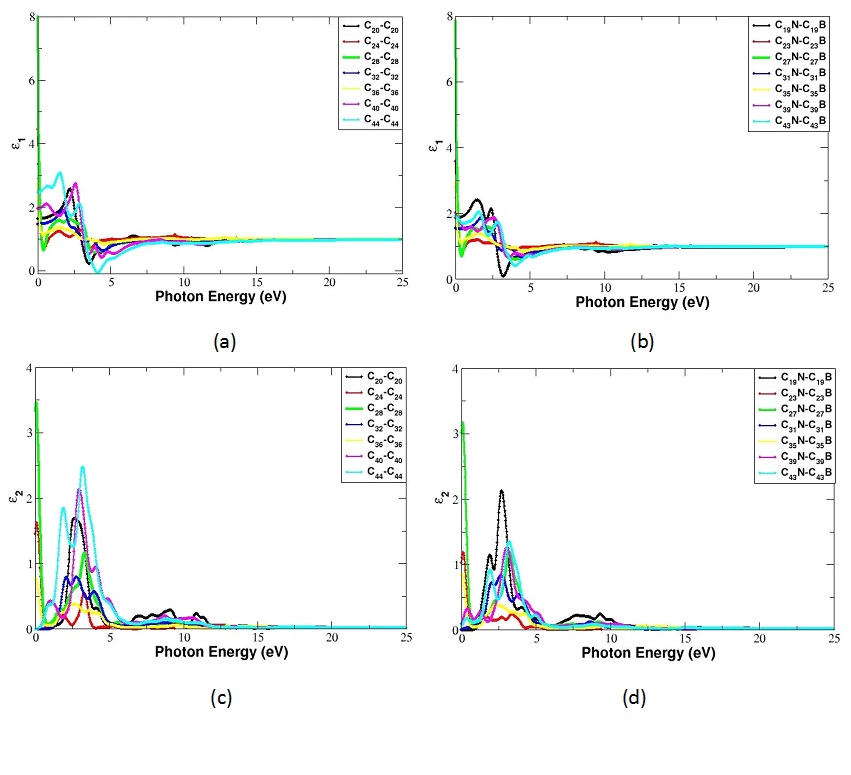}
\caption{Real ($\epsilon_{1}$; a,b) and imaginary ($\epsilon_{2}$;c,d) parts of dielectric function for Pure (a,c) and Doped (b,d) dimers.}
\label{fig:epsilon}
\end{figure}

The optical properties of the pristine and doped dimers have been investigated by application of an average electric field. The frequency dependent dielectric function is a key parameter, which provides information about nature of electromagnetic interaction with system\cite{paul,yang}.
Figure \ref{fig:epsilon} show the real and imaginary parts of dielectric function for pure (Figure \ref{fig:epsilon} a,c) and doped (Figure \ref{fig:epsilon} b,d) small fullerene dimers respectively.
The magnitude of static dielectric function for pure $C_{20}$ dimer is 1.64 which increases to 1.86 for doped dimer. In a similar way, the static dielectric function increases for $C_{32}$, $C_{36}$ and $C_{40}$ dimers on doping with N and B atoms, whereas for $C_{24}$, $C_{28}$ and $C_{44}$ dimers its magnitude decrease after doping. In addition to this, among all the considered pure fullerene dimers, the highest value of static dielectric function is 8.32 for $C_{28}$ dimer which reduces to 7.84 after their doping, while the lowest value is for $C_{32}$ dimer (1.47) which increases to 1.54 after its doping. In an optical spectrum, the first absorption peak gives the optical transition threshold, which is related to optical gap i.e. the energy gap between highest occupied and lowest unoccupied energy states of fullerenes\cite{paul}. The spectrum of imaginary part of dielectric function (Figure \ref{fig:epsilon} c and \ref{fig:epsilon} d) represents the optical gaps of pure and doped $C_{20}$ and $C_{32}$ dimers, which lies in the visible region whereas for other dimers the optical gaps shift to lower energies (higher wavelengths) in the infrared region. Among pure dimers, $C_{20}$ dimer has highest optical gap (2.65 eV) whereas $C_{36}$ dimer has lowest optical gap (0.06 eV). After their doping, $C_{31}$N-$C_{31}$B dimer has the highest optical gap (2.03 eV) while $C_{35}$N-$C_{35}$B dimer has lowest optical gap (0.04 eV). For $C_{20}$, $C_{36}$, $C_{40}$ and $C_{44}$ doped dimers, the optical gaps are red shifted w.r.t. their pure counterparts, while for $C_{24}$, $C_{28}$ and $C_{32}$ the optical gaps shift slightly towards lower wavelength. The presence of optical gaps in all considered dimers show that these fullerene dimers are optical semiconductors\cite{paul, Dpaul}. The optical gaps are consistent with HOMO-LUMO gaps which also confirm the semiconducting nature of the dimer structures. The doped $C_{32}$ dimer, which has lowest dielectric function and highest optical gap among all the other doped dimers can be used in short-wavelength optoelectronic device application.

\begin{figure}
\centering
\includegraphics[width=6in]{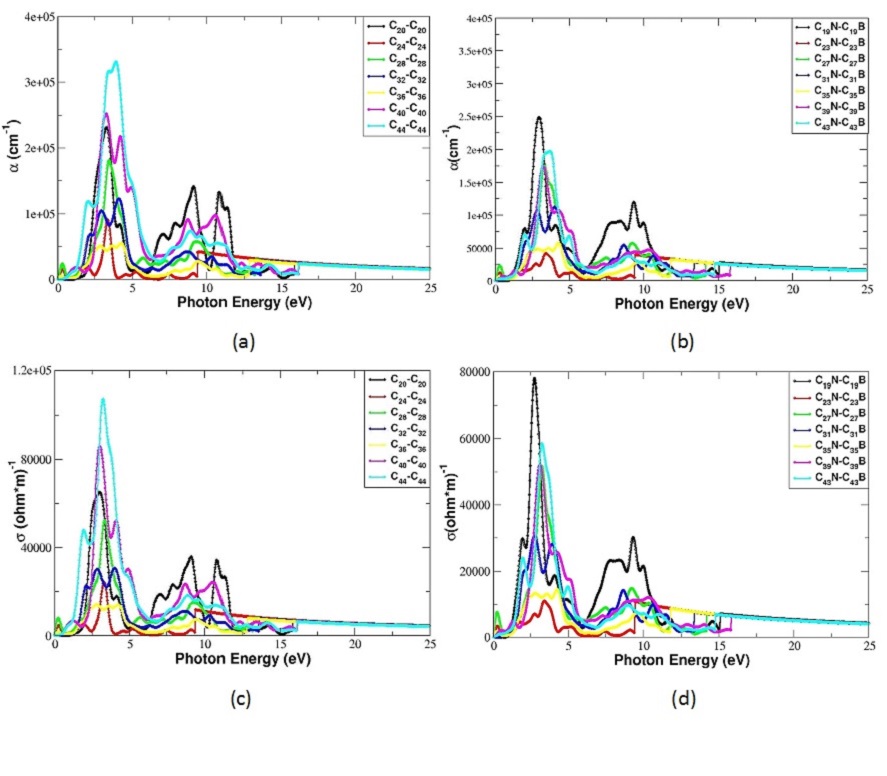}
\caption{Variation in absorption coefficient (a,b) and optical conductivity (c,d) with photon energy for Pure (a,c) and Doped (b,d) dimers.}
\label{fig:conductivity}
\end{figure}

Figure \ref{fig:conductivity} a,b represents the energy dependence of absorption coefficient for pure and doped dimer structures.  It is clear from the spectrum that absorption starts in infrared region for all the dimers and after their doping the onsets of spectrum shift towards longer wavelengths. For pure dimers, the highest absorption peak falls in Ultraviolet (UV) region. The highest peaks of absorption spectrum are redshifted for doped $C_{20}$, $C_{28}$, $C_{32}$, $C_{36}$, $C_{40}$ and $C_{44}$ dimers, whereas for doped $C_{24}$ the highest peak is blueshifted w.r.t. its pure dimer. In case of $C_{20}$ dimer, the doping of dimer with (N,B) leads to shifting of the highest absorption peak from UV to visible region. For all the dimers the strongest peak occur within the energy range 1.5-12 eV, which falls in UV and visible region indicating the maximum absorption. The optical conductivity have also been plotted for pure and doped dimers in Figure \ref{fig:conductivity} c,d. 
The optical conductivities for all the pure and doped dimers mimic the trend as seen for their absorption coefficients (Figure \ref{fig:conductivity} a,b) and the only difference lie between the amplitude of two spectrum. The peak height of absorption coefficients exceed the peak height in optical conductivity.

\begin{figure}
\centering
\includegraphics[width=6in]{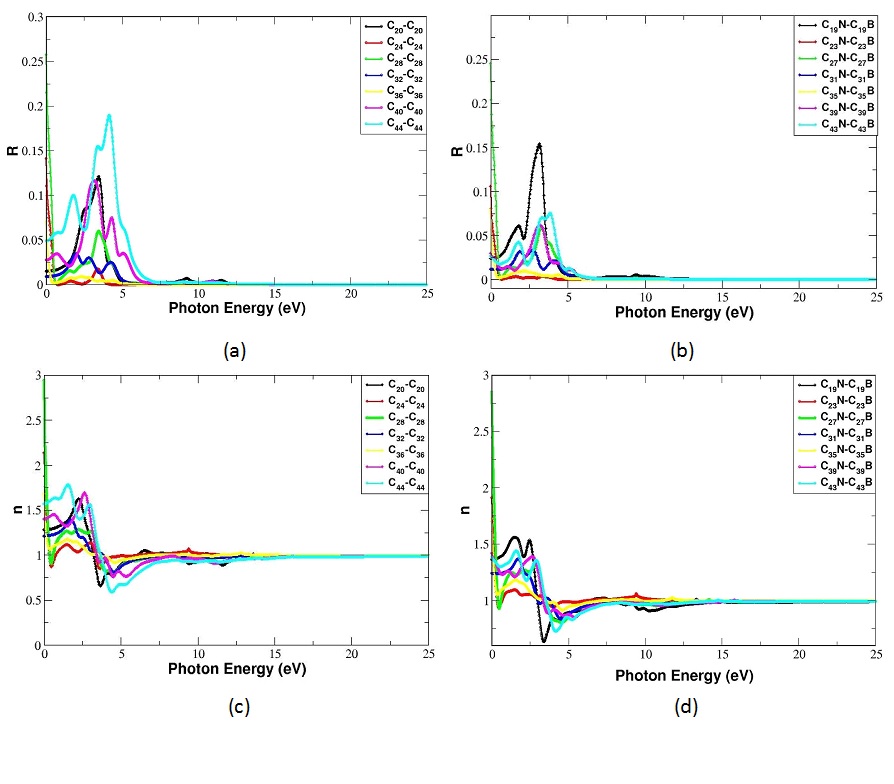}
\caption{Reflectivity (a,b) and Refractive Index (c,d) for Pure (a,c) and Doped (b,d) dimers.}
\label{fig:Rindex}
\end{figure}

The values of reflectivity and refractive index are shown in Figure \ref{fig:Rindex}. The highest refractive index for pure $C_{20}$, $C_{32}$, $C_{40}$ and $C_{44}$ dimers occur at photon energies of 2.29 eV, 1.8 eV, 2.68 eV and 1.61 eV, which fall in visible region and have magnitude 1.63, 1.39, 1.69 and 1.78 respectively. The highest peak of reflectivity shows that the system reflects photon of that particular energy at which highest peak occurs. The doping of these dimers shifts the highest peak towards longer wavelength between energy range 1.52-2.65 eV. Decrease in their highest refractive index also occur with doping. For $C_{24}$, $C_{28}$ and $C_{36}$ pure dimers, the highest refractive index lies in IR region with magnitudes of 2, 2.72 and 1.59 respectively. The doping of these dimers results in the decrease of magnitude of refractive index and hence transparency of system also increases. 
The pure and doped dimers show low values of reflectivity as compared to refractive index and vice versa. This particular property is used for optoelectronic device applications\cite{paul}.

\section{Conclusions}
The substitutional doping by N and B of symmetrical fullerene dimers ($C_{20}$, $C_{24}$, $C_{28}$, $C_{32}$, $C_{36}$, $C_{40}$ and $C_{44}$) has been investigated using DFT calculations to study their electronic and magnetic properties. The interaction energy values show that all the doped dimers are energetically favourable and are likely to be formed experimentally. The chemical hardness factor infers that the doped dimer structures have higher chemical reactivity as compared to other B and N doped carbon materials like sumanene. There is an opening of cage structure at junction of two cages with substitutional doping with N and B in $C_{20}$ and $C_{40}$ dimers. The total MM of pure magnetic dimers decrease with doping, while for non-magnetic pure dimers, it remains the same. All the magnetic dimers considered have shown ferromagnetic nature of magnetic interaction between two cages. For magnetic dimers, the HOMO-LUMO gaps for spin up and spin down states have finite energy difference which results in their magnetic behaviour, whereas the non-magnetic dimers have isoenergetic spin up and spin down states. The electronic DOS depicts that the conductivity of the pure dimers increase with doping due to redistribution and polarization of charge. Muliken charge analysis reveals that there is significant charge transfer of order 0.5 e from N to B atom. The surrounding C atoms donate 0.2 e to B due to their higher electronegativity. The optical properties of dimers calculated in the presence of an applied electric field show them behaving as optical semiconductors. The lowest dielectric function and highest optical gap of $C_{32}$ dimer can be used in short-wavelength region. The doping of $C_{20}$ dimer with B and N lead to the tuning of the absorption coefficient from UV to visible region. For all the dimers, the maximum absorption takes place in UV and visible region. The doping of dimers results in decrease of refractive index and hence transparency of system increases.
The tunable electronic, magnetic and optical properties of these dimers facilitate them as a good conducting as well as magnetic materials, which can have potential applications in both spintronics and electronics.

\section*{Acknowledgements}

Authors are thankful to Siesta group for providing their computational code. Authors are also grateful to UGC (University Grant Commission)-New Delhi, India for their financial support.

\newpage
\bibliographystyle{unsrt}

\begin{thebibliography} {}

\bibitem{kroto} H.W. Kroto, J.R. Heath, S.C.O'Brien, R.F. Curl and R.E. Smalley, $C_{60}$: Buckminsterfullerene Nature \textbf{1985} 318, 162-163.
\bibitem{Wkroto} H.W. Kroto, The stability of the fullerenes $C_{n}$, with n = 24, 28, 32, 36, 50, 60 and 70, Nature \textbf{1987} 329, 529-531.
\bibitem{kratschmer} W. Kratschmer, L.D. Lamb, K. Foristopoulos and D.R. Huffman, Solid $C_{60}$: a new form of carbon, Nature \textbf{1990} 347, 354-358.
\bibitem{diederich} F.Diederich, R. Ettl, Y. Rubin, R.L. Whetten, R. Beck, M.Alvarez, S. Anz et.al, The higher fullerenes: Isolation and characterization of $C_{76}$, $C_{84}$, $C_{90}$, $C_{94}$ and $C_{70}$O, an oxide of $D_{5h}-C_{70}$, Science \textbf{1991} 252, 548-551.
\bibitem{piskoti} C. Piskoti, J. Yarger and A. Zettl, $C_{36}$, a new carbon solid, Nature \textbf{1998} 393, 771-774.
\bibitem{an} Y.P. An, C.L. Yang, M.S. Wang, X.G. Ma and D.H. Wang, First-principles study of structure and quantum transport properties of $C_{20}$ fullerene, J. Chem. Phys. \textbf{2009} 131, 024311(1-6).
\bibitem{celaya} C.A. Celaya, J. Muniz, L.E. Sansores, New nanostructures of carbon: Quasi fullerenes $C_{n-q}$ (n=20, 42, 48, 60), Comp. and Theo. Chem. \textbf{2017} 1117, 20-29.
\bibitem{breda} N. Breda, R.A. Broglia, G. Colo, G. Onida, D. Provasi and E. Vigezzi, $C_{28}$: A possible room temperature organic superconductor, Phys. Rev. B \textbf{2000} 62, 130-133.
\bibitem{prinzbach} H. Prinzbach, A. Weiler, P. Landenberger, F. Wahl, J. Worth, L.T. Scott, M. Gelmont, D. Olevano and B.v. Issendorff, Gas-phase production and photoelectron spectroscopy of smallest fullerene, $C_{20}$, Nature \textbf{2000} 407, 60-63.
\bibitem{podlivaev} A.I. Podlivaev, K.P. Katin, D.A. Lovanov, L.A. Openov, Specific features of the stone-wales transformation in $C_{20}$ and $C_{36}$ fullerenes, Phys. Solid State \textbf{2011} 53, 215-220.
\bibitem{Apodlivaev} A.I.Podlivaev and K.P. Katin, On the dependence of the lifetime of an atomic cluster on the intensity of its heat exchange with the environment, JETP Lett. \textbf{2010} 92, 52-56.
\bibitem{tendero} S.G. Tendero, G. Sanchez, M. Alcami and F. Martin, Ionization potentials and dissociation energies of neutral, singly and doubly charged $C_{n}$ fullerenes from n = 20 to 70, Int. J. Mass. Spec. \textbf{2006} 252, 133-141.
\bibitem{bernstein} L.S. Bernstein, R.M. Shroll, D.K. Lynch and F.O. Clark, A small fullerene $C_{24}$ may be the carrier of 11.2${\mu}$m UIR, The Astrophysical Journal \textbf{2017} 836, 229(1-11).
\bibitem{sedel} O.V. Sedel'nikova, L.G. Bulusheva and A.V. Okotrub, Influence of defects in the carbon network on the static polarizability of fullerenes, Phys. Solid State, \textbf{2009} 51, 863-869. 
\bibitem{pradeep} T. Pradeep, V. Vijayakrishnan, A.K. Santra and C.N.R. Rao, Interaction of nitrogen with fullerenes: nitrogen derivatives of $C_{60}$ and $C_{70}$, J. Phys. Chem. \textbf{1991} 95, 10564-10565.
\bibitem{Ryu} R. Yu, M. Zhan, D. Cheng, S. Yang, Z. Liu and L. Zheng, Simultaneous synthesis of carbon nanotubes and nitrogen doped fullerenes in nitrogen atmosphere, J. Phys. Chem. \textbf{1995} 99 1818-1819.
\bibitem{otero} G. Otero, G. Biddau, C. Sanchez-Sanchez, R. Caillard, M.F.Lopez, C.Rogero, F.J. Palomares et.al, Fullerenes from aromatic precursors by surface-catalysed cyclohydrogenation, Nature \textbf{2008} 454, 865-868.
\bibitem{strout} D.L. Strout, Structure and stability of boron nitrides: The crossover between rings and cages, J. Phys. Chem. A, \textbf{2001} 105, 261-263.
\bibitem{rad} A.S. Rad and K. Ayub, Nonlinear optical and electronic properties of Cr-, Ni- and Ti- substituted $C_{20}$ fullerenes: A quantum-chemical study, Mater. Res. Bull. \textbf{2018} 97, 399-404.
\bibitem{amiri} S.S. Amiri, M. Koohi and B. Mirza, Characterizations of B and N heteroatoms as substitutional doping on structure, stability and aromaticity of novel heterofullerenes evolved from the smallest fullerene cage $C_{20}$: a density functional theory perspective, J. Phys. Org. Chem. \textbf{2016} 29, 514-522.
\bibitem{katin} K.P. Katin and M.M. Maslov, Stone-wales defect in nitrogen doped $C_{20}$ fullerenes:Insight from ab-initio calculations, Physica E \textbf{2018} 96, 6-10.
\bibitem{fowler} P.W. Fowler, T. Heine, K.M. Rogers, J.P.B. Sandall, G. Seifert and F. Zerbetto, $C_{36}$, a hexavalent building block for fullerene compounds and solids, Chem. Phys. Lett. \textbf{1999} 300, 369-378.
\bibitem{maryam} M. Anafcheh and R. Ghafouri, Theoretical studies on one-dimensional polymers constructed from BN-substituted $C_{36}$ fullerene, Compu. Theo. Chem. \textbf{2013} 1017, 1-6.
\bibitem{viani} L. Viani and M.C. dos Santos, Comparative study of lower fullerenes doped with boron and nitrogen, Solid State Commun. \textbf{2006} 138, 498-501.
\bibitem{kaur} S. Kaur, A. Sharma, H. Sharma and I. Mudahar, Structural and magnetic properties of small symmetrical and asymmetrical sized fullerene dimers, Mater. Res. Exp. \textbf{2018} 5, 016105(1-15).
\bibitem{GWwang} G.W. Wang, K. Komatsu, Y. Murata and M. Shiro, Synthesis and x-ray structure of dumb-shell shaped $C_{120}$, Nature \textbf{1997} 387, 583-586.
\bibitem{Nkaur} N. Kaur, S. Gupta, K. Dharamvir and V.K. Jindal, The formation of dimerized molecules of $C_{60}$ and their solids, Carbon, \textbf{2008} 46, 349-358.
\bibitem{Kkomatsu} K. Komatsu, G.W. Wang, Y. Murata, T. Tanaka, K. Fujiwara, K. Yamamoto and M. Saunders, Mechanochemical synthesis and characterization of fullerene dimer, J. Org. Chem. \textbf{1998} 63, 9358-9366.
\bibitem{sabirov} D.Sh. Sabirov, Polarizability of $C_{60}$ fullerene dimer and oligomers: the unexpected enhancement and its use for rational design of fullerene-based nanostructures with adjustable properties, RSC Adv. \textbf{2013} 3, 19430-19439.
\bibitem{denis} D.Sh. Sabirov, A.O. Terentyev and R.G. Bulgakov, Polarizability of fullerene [2+2] dimers: a DFT study, Phys. Chem. Chem. Phys. \textbf{2014} 16, 14594-14600.
\bibitem{swart} M. Swart and P.Th. van Duijnen, Rapid determination of polarizability exaltation in fullerene-based nanostructures, J. Mater. Chem. C, \textbf{2015} 3, 23-25.
\bibitem{alina} A.A. Tukhbatullina, I.S. Shepelevich and D.Sh. Sabirov, Exaltation of polarizability as a common property of fullerene dimers with diverse intercage bridges, Fullerenes, Nanotubes and Carbon Nanostructures, \textbf{2018} 26, 661-666.
\bibitem{bhattacharya} B. Bhattacharya, NB Singh, R. Mondal and U. Sarkar, Electronic and optical properties of pristine and boron-nitrogen doped graphyne nanotubes, Phys Chem Chem Phys \textbf{2015} 17, 19325-19341.
\bibitem{NBsingh} B. Bhattacharya, N.B. Singh and U. Sarkar, Pristine and BN doped graphyne derivatives for UV light protection, Int J Quantum Chem.\textbf{2015} 115, 820-829.
\bibitem{ching} W.Y. Ching, M.Z. Huang, Y.N. Xu, W.G. Harter and F.T. Chan, First principles calculation of optical properties of $C_{60}$ in the fcc lattice, Phys. Rev. Lett. \textbf{1991} 67, 2045-2048.
\bibitem{paul} D. Paul, J. Deb, B. Bhattacharya and U. Sarkar, Electronic and optical properties of $C_{24}$, $C_{12}X_{6}Y_{6}$, and $X_{12}Y_{12}$ (X = B, Al and Y = N, P), J Mol. Model. \textbf{2018} 24, 204(1-13).
\bibitem{Dpaul} D. Paul, B. Bhattacharya, J. Deb, U. Sarkar, Optical properties of $C_{28}$ fullerene cage: a DFT study, AIP Conf. Proc. \textbf{2018} 1953, 030236(1-3).\bibitem{junquera} J. Junquera, O. Paz, D.S. Portal, and E. Artacho, Numerical atomic orbitals for linear scale calculations, Phys. Rev. B, \textbf{2001} 64, 235111(1-9).
\bibitem{perdew} P. Perdew, K. Burke and M. Ernzerhof, Generalized gradient approximation made simple, Phys. Rev. Lett, \textbf{1996} 77, 3865-3868.
\bibitem{kleinman} L. Kleinman, D.M. Bylander, Efficacious form for model pseudopotentials, Phys. Rev. Lett. \textbf{1982} 48, 1425-1428.
\bibitem{sankey} O.F. Sankey, D.J. Niklewski, Ab initio multicenter tight binding model for molecular dynamics simulations and other applications in covalent systems, Phys. Rev. B, Condens. Matter, \textbf{1989} 40, 3979-3995.
\bibitem{feyereisen} M. Feyereisen, M. Gutowski and J. Simons, Relative stabilities of fullerene, cumulene and polyacetylene structures for Cn:
n=18–60, J. Chem. Phys. \textbf{1992} 96, 2926–2932.
\bibitem{garg} I. Garg, H. Sharma, N. Kapila, K. Dharamvir and V.K. Jindal, Transition metal induced magnetism in small fullerenes (Cn for $n\leq36$) Nanoscale, \textbf{2011} 3, 217–224.
\bibitem{hitesh} H. Sharma, I. Garg, K. Dharamvir and V.K. Jindal, Structural, electronic and vibrational properties of $C_{60-n}N_n$ (n=1–12) J. Phys.
Chem. A, \textbf{2009} 113, 9002–9013.
\bibitem{amrish} A. Sharma, S. Kaur, H. Sharma, I. Mudahar, Electronic and magnetic properties of small fullerene carbon nanobuds: A DFT Study, Mater. Res. Express, \textbf{2018} 5, 065032.
\bibitem{skaur} S. Kaur, H. Sharma and I. Mudahar, Substitutional doping of asymmetrical small fullerene dimers, Adv. Sci. Lett. \textbf{2018} 24, 888-892.
\bibitem{asharma} A. Sharma, H. Sharma and I. Mudahar, A first principle study on $C_{20}$ and $C_{40}$ carbon nanobuds, Adv. Sci. Lett. \textbf{2018} 24, 790-795.
\bibitem{Anafcheh} M. Anafcheh and R. Ghafouri, Exploring the electronic and magnetic properties of $C_{60}$ fullerene dimers with ladderane-like
hexagonal bridges, Comput. Theor. Chem. \textbf{2012} 1000, 85–91.
\bibitem{Ma} F. Ma, Z.R. Li, Z.J. Zhou, D. Wu, Y. Li, Y.F. Wang and Z.S. Li, Modulated nonlinear optical responses and charge transfer transition in endohedral fullerene dimers Na@$C_{60}C_{60}$@F with n-fold covalent bond (n=1,2,5 and 6) and long range ion bond, J. Phys. Chem. C \textbf{2010} 114, 11242-11247.
\bibitem{segura} J.L. Segura and N. Martin, [60]Fullerene dimers, Chem. Soc. Rev. \textbf{2000} 29, 13-25.
\bibitem{adel} A.R. Vanani and S. Mehrdoust, Effect of boron doping in sumanene frame toward hydrogen physisorption: A theoretical study, Int J Hydrogen Energy, \textbf{2016} 41, 15254-15265.
\bibitem{vanani} A. R. Vanani and F. Shamsali, Influence of nitrogen doping in sumanene framework toward hydrogen storage: A Computational study, J Mol Graph Model, \textbf{2017} 76, 475-487.
\bibitem{yang} L.M. Yang, P. Ravindran, P. Vajeeston and M. Tilset, Ab-initio investigations on crystal structure , formation enthalpy, electronic structure, chemical bonding and optical properties of experimentally synthesized isoreticular metal-organic framework-10 and its analogues:M-IRMOF-10 (M=Zn, Cd, Be, Mg, Ca, Sr and Ba), RSC Adv. \textbf{2012} 2, 1618-1631.
  
\end{thebibliography}

\newpage
\begin{table}
\caption{Interaction energies ($E_{int}$), chemical hardness ($\eta$) and electronegativity ($\chi_{m}$) of all possible isomers of symmetrical small fullerene dimers}
\begin{center}
\begin{tabular}{|c|c|c|c|c|c|c|}
\hline
Dimer & $E_{int}$ [Ref. 25] (eV) & Dimer & Isomers & $E_{int}$ (eV) & $\eta$ (eV) & $\chi_{m}$ (eV) \\
\hline
$C_{20}-C_{20}$[2+2] & -4.35 & $C_{19}N-C_{19}B$ & (9,39) & -4.59 & 0.65 & 3.50 \\
& & & (12,40) & -4.43 & 0.64 & 3.50 \\
\hline
$C_{24}-C_{24}$[6+6] & -5.18 & $C_{23}N-C_{23}B$ & (22,29) & -4.07 & 0.09 & 3.97 \\
& & & (5,46) & -3.65 & 0.08 & 3.97 \\
& & & (23,32) & -3.57 & 0.08 & 3.98 \\
& & & (8,47) & -3.47 & 0.09 & 3.97 \\
& & & (4,28) & -2.97 & 0.09 & 3.98 \\
& & & (24,48) & -2.64 & 0.08 & 4.00 \\
\hline
$C_{28}-C_{28}$[5+5] & -2.01 & $C_{27}N-C_{27}B$ & (13,34) & -3.00 & 0.05 & 3.92 \\
& & & (19,31) & -2.19 & 0.02 & 3.92 \\
& & & (22,36) & -1.93 & 0.05 & 3.91 \\
& & & (14,32) & -1.84 & 0.05 & 3.92 \\
& & & (23,52) & -1.66 & 0.03 & 3.93 \\
\hline
$C_{32}-C_{32}$[2+2] & -1.55 & $C_{31}N-C_{31}B$ & (19,43) & -3.00 & 0.15 & 3.79 \\
& & & (28,40) & -1.71 & 0.17 & 3.86 \\
\hline
$C_{36}-C_{36}$[2+2] & -1.57 & $C_{35}N-C_{35}B$ & (8,46) & -1.34 & 0.03 & 3.70 \\
& & & (13,54) & -0.60 & 0.03 & 3.69 \\
\hline
$C_{40}-C_{40}$[2+2] & -0.72 & $C_{39}N-C_{39}B$ & (25,65) & -1.74 & 0.06 & 3.74 \\
& & & (27,67) & -1.74 & 0.06 & 3.74 \\
\hline
$C_{44}-C_{44}$[2+2] & -0.34 & $C_{43}N-C_{43}B$ & (18,56) & -1.38 & 0.08 & 3.99 \\
& & & (20,54) & -0.86 & 0.08 & 3.99 \\
\hline
\end{tabular}
\end{center}
\end{table}

\begin{table}
\caption{Muliken charge analysis for most stable structures of doped symmetrically sized fullerene dimers. + and - sign indicates loose and gain in charge, respectively.}
\begin{center}
\begin{tabular}{|c|c|c|}
\hline
Dimer & N$^{+}$ & B$^{-}$ \\
\hline
C$_{19}$N-C$_{19}$B & 0.560 & 0.747 \\
C$_{23}$N-C$_{23}$B & 0.398 & 0.648 \\
C$_{27}$N-C$_{27}$B & 0.407 & 0.663 \\
C$_{31}$N-C$_{31}$B & 0.448 & 0.746 \\
C$_{35}$N-C$_{35}$B & 0.460 & 0.715 \\
C$_{39}$N-C$_{39}$B & 0.511 & 0.851 \\
C$_{43}$N-C$_{43}$B & 0.450 & 0.762 \\
\hline
\end{tabular}
\end{center}
\end{table}

\end{document}